\documentclass[reprint,showpacs,preprintnumbers,amsmath,amssymb, rsi,prl,superscriptaddress]{revtex4-1}
\usepackage{graphicx,wasysym}
\usepackage{dcolumn}
\usepackage{bm}
\usepackage{natbib}
\usepackage{tabulary}
\usepackage{hyperref}

\usepackage{color}
\usepackage[dvipsnames]{xcolor}
\usepackage[normalem]{ulem}

\newcommand{\kB}{k_\mathrm{B}}

\usepackage{bbold}

\usepackage{lipsum}

\begin{document}

\preprint{1}

\title{Solid-state ionic rectification in perovskite nanowire heterostructures}
\author{Qiao Kong}
\thanks{These authors contributed equally.}
 \affiliation{Department of Chemistry, University of California, Berkeley}
\author{Amael Obliger}
\thanks{These authors contributed equally.}
 \affiliation{Laboratoire des Fluides complexes et leurs Réservoirs, UMR 5150, Université de Pau et des Pays de l'Adour, E2S-UPPA/CNRS/TOTAL, Pau, France}
 \author{Minliang Lai}
 \affiliation{Department of Chemistry, University of California, Berkeley}
\author{Mengyu Gao}
 \affiliation{Department of Materials Science and Engineering, University of California, Berkeley}
 \affiliation{Materials Science Division, Lawrence Berkeley National Laboratory}
\author{David T. Limmer}
 \email{dlimmer@berkeley.edu}
 \affiliation{Department of Chemistry, University of California, Berkeley}
\affiliation{Materials Science Division, Lawrence Berkeley National Laboratory}
\affiliation{Chemical Science Division, Lawrence Berkeley National Laboratory}
\affiliation{Kavli Energy NanoScience Institute, Berkeley, California, Berkeley}
\author{Peidong Yang}
 \email{p_yang@berkeley.edu}
 \affiliation{Department of Chemistry, University of California, Berkeley}
\affiliation{Department of Materials Science and Engineering, University of California, Berkeley}
\affiliation{Materials Science Division, Lawrence Berkeley National Laboratory}
\affiliation{Kavli Energy NanoScience Institute, Berkeley, California, Berkeley}

\date{\today}

\maketitle

{\bf Halide perovskites have attracted increasing research attention regarding their outstanding optoelectronic applications. Owing to its low activation energy, ion migration is implicated in the long-term stability and many unusual transport behaviors of halide perovskite devices. However, precise control of the ionic transport in halide perovskite crystals remains challenging. Here we visualized and quantified the electric-field-induced halide ion migration in an axial CsPbBr$_3$-CsPbCl$_3$ nanowire heterostructure and demonstrated a solid-state ionic rectification, which is due to the non-uniform distribution of the ionic vacancies in the nanowire that results from a competition between electrical screening and their creation and destruction at the electrode interface. The asymmetric heterostructure characteristics add an additional knob to the ion-movement manipulation in the design of advanced ionic circuits with halide perovskites as building blocks.
}

As a new class of semiconductor materials, halide perovskites have been intensively researched in connection with a rich variety of photovoltaic and light-emitting properties \cite{brittman2015expanding,kojima2009organometal,green2014emergence,lee2012efficient,xing2014low,zhu2015lead,cho2015overcoming,lin2018perovskite,cao2018perovskite,wang2016perovskite}. Studies in halide perovskite nanomaterials are extended to the creation of functional heterostructures, which is important for generating unique material characteristics and constructing advanced optoelectronic devices\cite{dou2017spatially,wang2017epitaxial,kong2018phase}. The diversity of halide perovskites enables the facile solution-processed materials integration in single nanostructures, allowing for the investigation of charge carrier dynamics across the heterointerfaces. Apart from their extraordinary optoelectronic performance, the dynamic and reconfigurable halide perovskite lattice exhibits additional perspective of designing and controlling the lattice ion migration in the formed heterostructures, giving rise to the ionic transport manipulation in solid-state systems\cite{pan2018visualization,nedelcu2015fast,tress2017metal,mizusaki1983ionic,xiao2015giant,game2017ions,calado2016evidence}. 

Here, we investigate halide ion migration and interdiffusion under applied voltage across a CsPbBr$_3$-CsPbCl$_3$ nanowire heterostructure formed on pre-patterned parallel Au electrodes via a previously demonstrated anion exchange process\cite{dou2017spatially,lai2018intrinsic}. The relationship between halide composition in lead halide perovskites and their peak photoluminescence (PL) wavelength enables the visualization and quantification of local halide movement throughout the nanowire. Using confocal PL, we observe that the rate of halide migration upon applied voltage depends on the polarity of the bias across the heterojunction and subsequently the nanowires are able to rectify ionic motion. Such behavior is similar to that widely observed in nanofluidic devices made from electrolyte solutions\cite{lin2018voltage,cheng2009ionic,yan2009nanofluidic,fan2005polarity}.  We develop a theoretical model based on the coupled reaction-drift-diffusion dynamics of halide ions, their charged vacancies, and the electrostatic potential, to explain the observed nonlinear ionic dynamics. We find that ionic rectification in solid-state halide perovskites arises from the non-uniform, quasi-steady state distribution of halide vacancies in the heterostructure, which in turn results from an interplay between screening the applied voltage and their creation and annilation at the nanowire-electrode interface. The theoretical model is in excellent agreement with the experimental observations, offering opportunities for solid state engineering applications.

We prepared the CsPbBr$_3$-CsPbCl$_3$ nanowire heterostructure on a pre-patterned Au bottom contact and observed the halide ion migration under an electric field applied across the heterostructure interface. Specifically, CsPbBr$_3$ nanowires were synthesized via our previously reported solution-processed method, exhibiting diameters with hundreds of nanometers and lengths up to 20 micrometers (Fig. S1). X-ray power diffraction (XRD) characterization showed consistent features with standard orthorhombic perovskite phase patterns (Fig. S2). These nanowires were transferred onto the parallel Au electrodes patterned on the SiO$_2$/Si substrate, which was subsequently covered by poly(methyl methacrylate) (PMMA). Using electron-beam lithography, we exposed the area with the nanowire underneath where the anion-exchange from Br to Cl was conducted (see Supplementary Materials). The CsPbBr$_3$-CsPbCl$_3$ heterostructure with well-defined axial nanowire geometry is shown by the scanning electron microscopy (SEM) image in Figure 1a. The as-prepared heterostructure nanowire featured green-blue two-color PL emission centered at 526 and 420 nm, indicating a clear and sharp interface, as well as a completed conversion from the CsPbBr$_3$ to CsPbCl$_3$ (Figure 1b). A characteristic snapshot of the molecular structure of the interface is shown in Figure 1c. 

\begin{figure}
\begin{center}
\includegraphics[width=8.5cm]{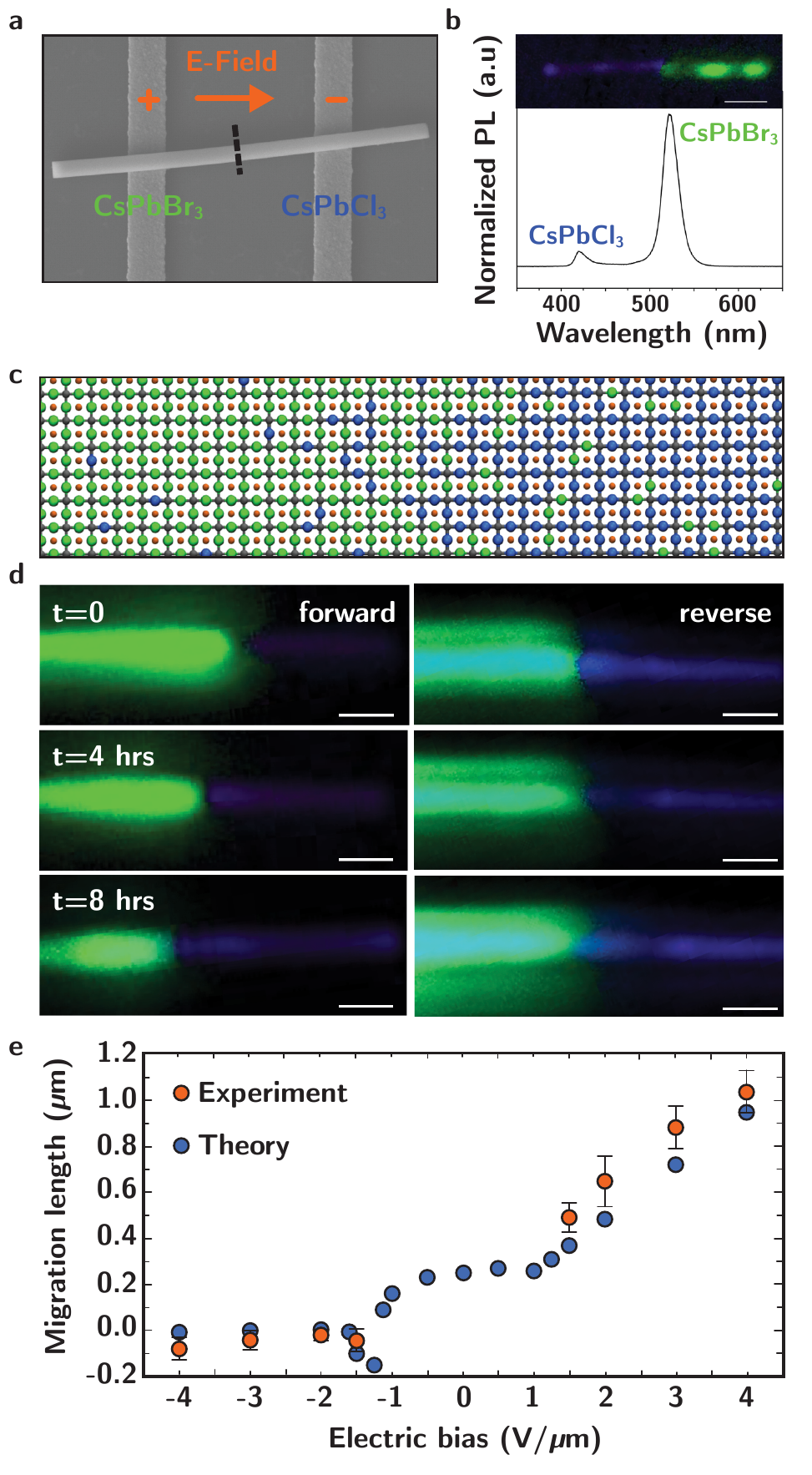}
\caption{1 Schematic illustration and characterization of electric-field-induced halide ion migration in CsPbBr$_3$-CsPbCl$_3$ heterostructure nanowires. (a) Scanning electron microscopic (SEM) image of the as-prepared heterostructure nanowire after anion exchange reactions in the exposed area of the pristine CsPbBr$_3$ nanowire. Scale bar is 1 $\mu$m. (b) Confocal photoluminescence (PL) image and emission spectra of the as-prepared heterostructure nanowire. CsPbBr$_3$ segment exhibited green emission with the peak position around 526 nm and CsPbCl$_3$ segment showed blue emission at 420 nm. Scale bar is 2$\mu$m. (c) Atomistic snapshot of a CsPbBr$_3$-CsPbCl$_3$ heterojunction (lattice spacing ~5.8 $\mathrm{\AA}$. Green: Br, Blue: Cl, Red: Cs, Grey: Pb) with end bromide concentrations corresponding to the experimental ones. (d) Confocal PL characterization of single heterostructure nanowire under electric field intensity 4 V/$\mu$m up to 8 hours with forward and reverse biasing. Scale bar is 2 $\mu$m. (e) Halide ion migration distance in 4-hour biasing with different field intensities (negative ones correspond to reverse biasing). The ion migration distance is defined as the movement of the interface front toward the CsPbBr$_3$ segment under biasing conditions. The composition front location used to evaluate the migration length is defined as the position of its mid-point $(x_\mathrm{Br}=0.55)$.}
\label{Fi:1}
\end{center} 
\end{figure}
At room temperature and under dark conditions, voltage biases with different intensities and directions were applied across the heterostructure nanowire to drive the halide ion  migration and interdiffusion. Due to the low charge carrier concentrations in the Pb-based halide perovskites, the electrical current was negligible regardless of the electric field directions, ruling out Joule heating effects to ion movement (Fig. S3). Spatially resolved PL spectra were examined after biasing to correlate the peak emission wavelength to the halide composition. Figure 1d shows the comparison of PL emission change near the interface under 4 V/$\mu$m field intensity with the positive electrostatic potential applied on the CsPbBr$_3$ segment, defined as the forward direction, and with the reversed polarity such that the positive potential was applied to the CsPbCl$_3$ segment. The green emission segment became shorter as Cl- migrated towards CsPbBr$_3$ segment under the forward biasing. The time evolution of the emission wavelength along the nanowire indicated that the interface shifted proportionally with time and remained sharp (Fig. S4). However, no PL emission change was observed with bias in the opposite direction. Such asymmetric transport was consistently observed under a series of forward and reverse biasing intensities. Within the electric field range of our experiment (1.5 – 4 V/$\mu$m), increasing forward bias intensity at fixed observation time led to a greater interface shift as shown by the decreasing length of the green emission segment (Fig. S5). By contrast, the interface was stationary under reverse biases within the same range of biasing time and field intensities employed (Fig. S6). 

This distinctly asymmetric field-induced ionic transport phenomena under forward and reverse bias is the hallmark of a diode. In this case, the heterostruture is found to rectify halide ion migration and interdiffusion, in which halide ions migrate proportionally to the applied bias towards the CsPbBr$_3$ segment, while the reverse bias does not allow anion migration. We quantified the extent of interdiffusion along the nanowire using the correlation between the emission wavelength and the halide composition (Fig. S7). Specifically, we define a migration length as the distance the interface moves in the direction of the bromide rich segment at a set observation time and under fixed bias. As shown in Figure 1e, the rectification behavior is highlighted by the migration length after four hours versus the electric bias. Under the forward bias, we found the migration length is linearly proportional to the bias for the experimentally investigated intensities. By contrast under the reverse bias, the migration length is nearly independent of bias and close to zero. Comparing the largest forward and reverse biases considered, we found a rectification ratio, or ratio of the migration lengths at $\pm$ 4 V/$\mu$m, of nearly 20. 

\begin{figure}[b]
\begin{center}
\includegraphics[width=8.5cm]{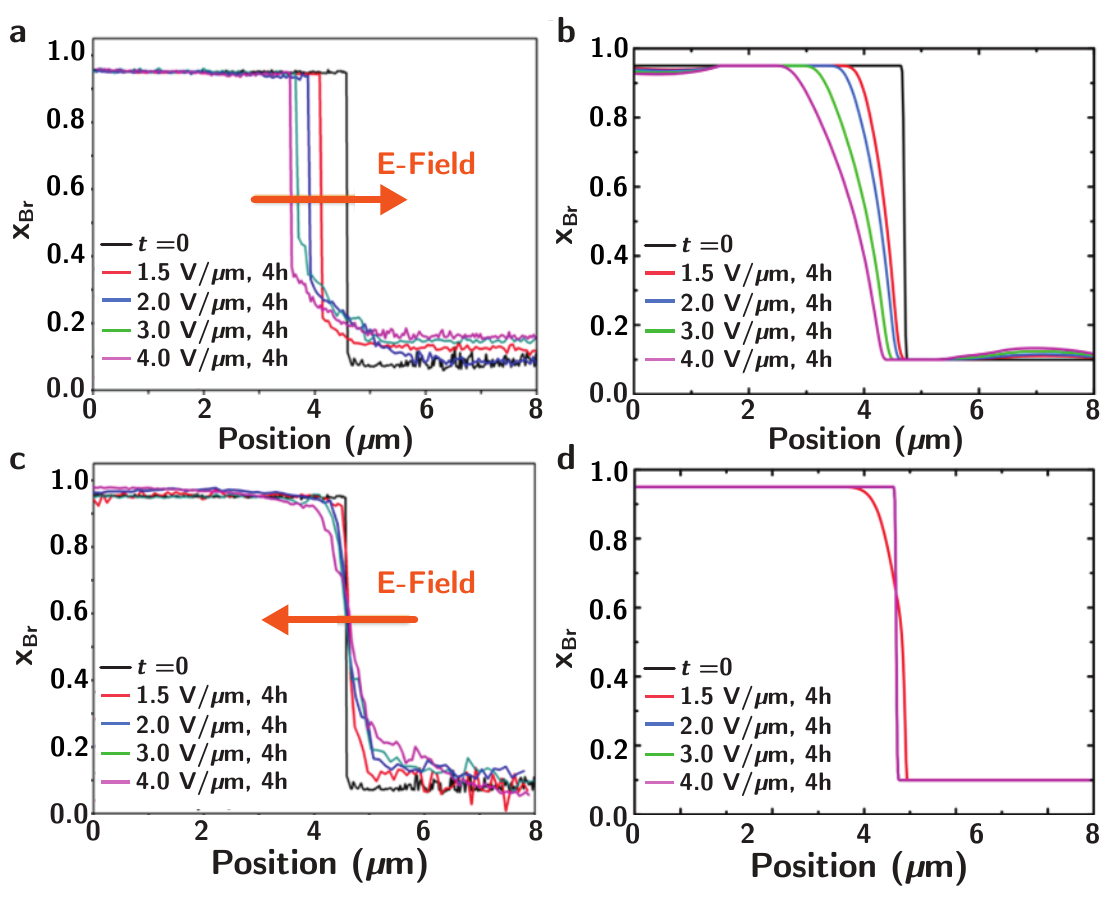}
\caption{Halide composition evolution along the CsPbBr$_3$-CsPbCl$_3$ heterostructure nanowire under 4-hour biasing with different electric field directions and intensities. (a) Halide composition profile in experiment (a) and as computed by the numerical model (b) along the heterostructure nanowire under forward biasing. Halide composition profile in experiment (c) and as computed by the numerical model (d) along the CsPbBr$_3$-CsPbCl$_3$ heterostructure nanowire under reverse biasing.}
\label{Fi:1}
\end{center} 
\end{figure}
As shown in Figure 2a, the estimated concentration of migrated Cl ions from the CsPbCl$_3$ into CsPbBr$_3$ segment is as high as 12\% of halide lattice sites within 4 hrs forward biasing at room temperature. This is a dramatic value compared to the field-induced lattice atom migration in many conventional inorganic solid-state materials. Similar ion migration behavior has also been reported in organic-inorganic hybrid counterparts\cite{zhang2016reversible,yuan2016electric}, exhibiting the universal dynamic lattice characteristics of the halide perovskite materials. A detailed analysis under the reverse bias in Figure 2c indicated the mixed Br and Cl compositions, suggesting a small amount of interdiffusion of halide anions across the CsPbBr$_3$/CsPbCl$_3$ interface. This was caused by the reverse bias as the control group without bias did not show any discernible emission wavelength change at room temperature (Fig. S8).  To eliminate the interference of the heterostructure formation and halide perovskite/Au electrode contact quality caused by the anion exchange process, we used an initial CsPbCl$_3$ nanowire as the starting material to create the same heterostructure. Consistent ion migration and interdiffusion were observed under forward and reverse biases, respectively (Fig. S9), confirming that this ionic rectifying behavior is an intrinsic property of the CsPbBr$_3$-CsPbCl$_3$ heterostructure. 

In order to explain the underlying mechanism of the observed solid-state rectification, we propose a coupled reaction-drift-diffusion model accounting simultaneously for the local dynamics of the halide anions, their charged vacancies, and the electrostatic potential. While in principle a set of coupled equations for each species in the heterojunction could be solved, previous work on lead halide perovskites allows us to make a number of simplifications. Halide transport is known to be vacancy mediated \cite{eames2015ionic,barboni2018thermodynamics},  with low Schottky pair binding energies\cite{lai2018intrinsic}. This allows us to disregard interstitial defects and model free vacancy diffusion.  Further, we can neglect the dynamics of the cations and their vacancies, since their characteristic diffusivities are orders of magnitude smaller than those of the anions due to their much larger activation barrier\cite{lai2018intrinsic}. Typical halide vacancy concentrations are $\sim$ 10$^{-7}$ per unit cell, implying a large separation of timescales between vacancy diffusion and halide diffusion, since the transport coefficients of the halides are proportional to the local vacancy concentration. We can thus expect that under an applied electric field, the vacancy distribution will establish a local equilibrium through which the halides will subsequently diffuse. The concentrations of vacancies are not large enough to completely screen the applied potential and at their initial concentrations their screening lengths are small relative to the size of the heterojunction. Thus, in the absence of vacancy sources, it is expected that little net halide transport would be observed. We find it sufficient to include linearized reversible reactions at the electrode that depend on the local halide composition and electric field to enable steady-state halide transport. This field and composition dependence in the vacancy reactions results in effective halide transport coefficients that depend on the applied field through the change in the quasi-static vacancy concentration. The heterojunction geometry introduces an initial asymmetry that when coupled to the electric field dependent transport results in ionic rectification. 

To quantify this basic physical mechanism and make direct contact with the experimental measurements, we first derive explicit forms for the reaction-drift-diffusion equations. Using previous molecular simulation results we can parameterize these equations directly, and solve them numerically using finite-element techniques\cite{courtier2018fast}. Specifically, we consider $x_\mathrm{V}, x_\mathrm{Br}$ and $x_\mathrm{Cl}$ for the halide vacancies, the bromide and the chloride ions. The conservation of lattice sites imply a normalization $x_\mathrm{Br}+x_\mathrm{Cl}+x_\mathrm{V}=1$, and enables us in practice to eliminate the chloride concentration from our dynamical description. Due to the small width of the nanowire, we treat the system along one dimensional and use the longitudinal coordinate $z$. The dynamics of the halide vacancies and bromides obey the following reaction-drift-diffusion equations\cite{belova2004analysis},
$$
\frac{\partial x_i}{\partial t} = \frac{\partial J_i}{\partial z} + S_i^\pm
$$
for $i=$Br, Cl and V, where $J_i$ is the flow of species $i$ and $S_i^\pm$ the source terms. The flows are generated in response to the the electric field $E$ and the affinities $X_a=-\kB T \partial \ln x_a/\partial z$ where $\kB T$ is Boltzmann’s constant times temperature, and can be expressed as
$$
J_\mathrm{Br} = L_{\mathrm{Br,Br}} \left ( X_{\mathrm{Br}} -X_{\mathrm{V}}- e E \right )+L_{\mathrm{Br,Cl}} \left ( X_{\mathrm{Cl}} -X_{\mathrm{V}}- e E \right )
$$
$$
J_\mathrm{Cl} = L_{\mathrm{Br,Cl}} \left ( X_{\mathrm{Br}} -X_{\mathrm{V}}- e E \right )+L_{\mathrm{Cl,Cl}} \left ( X_{\mathrm{Cl}} -X_{\mathrm{V}}- e E \right )
$$
$$
J_\mathrm{V} = -\left ( J_\mathrm{Br}+J_\mathrm{Cl} \right )
$$
with Onsager transport coefficients $L_{a,b}$ ($a,b=$Cl,Br) as derived by Moleko et al.\cite{moleko1989self} for vacancy mediate interdiffusion(SI). The source terms in the limit of low vacancy concentration,
$$
S_v^\pm =r_c^\pm \left ( 1- x_\mathrm{V}\right )-r_d^\pm  x_\mathrm{V}\, 
$$
$$
S_\mathrm{B}r^\pm =r_c^\pm x_\mathrm{B}+r_c^\pm \left (1- x_\mathrm{Br} \right ) x_\mathrm{V}
$$
are nonzero only at the interfaces with the electrodes  and introduce the creation and destruction rates of vacancies $r_c^\pm$ and $r_d^\pm$, respectively. The strength of these rates depends on the voltage difference applied at the electrodes, $r_i^\pm = \alpha_i^{\pm} |\Delta \psi|/2$ where $|\Delta \psi|$ is the applied potential and $\alpha_i^{\pm}$ the constant of proportionality between the rate and applied potential (see Supplementary Materials). These equations are solved together with Poisson’s equation for the electric field 
$$
\frac{\partial E}{\partial z}= \frac{\rho}{\epsilon}
$$
with $\epsilon=25\epsilon_0$ the electric permittivity of the lattice\cite{govinda2017behavior} and $\rho(z,t)=e c_s (x_\mathrm{V} (z,t)-x_\mathrm{V}^\mathrm{eq} (z,0))$ the electric density, where the cation vacancy concentration is fixed and equal to the initial anion vacancy concentration $x_\mathrm{v}^\mathrm{eq}(z,0)$ at equilibrium, $c_s=3/a^3$ is the anion lattice site density and $e$ is the elementary charge. No flow and zero electric field boundary conditions are enforced. Voltage values are imposed at the contact lines between the wire and the electrodes. The initial conditions are taken from the PL observations and are a step-like profile for the halide concentration, the electrostatic potential is zero everywhere and the vacancy concentration is assumed to start at equilibrium, $x_\mathrm{v}^\mathrm{eq}$, provided the known composition dependent free energies of formation\cite{lai2018intrinsic}. As we will discuss, this model is sufficient to reproduce the experimental rectification behavior quantitatively, as shown in Figure 1e.

Figure 2b shows that the forward biasing dynamics computed within the reaction-drift-diffusion model, which displays largely the same features as the experimental profiles shown in Figure 2a. A notable difference is that the anion composition front shifts while exhibiting a smoother shape as the bias intensity increases. Larger vacancy creation rates  allow for larger shifts, but also smooths the profile due to enhanced vacancy mediated  interdiffusion. We use the data at $\pm$ 2 V/$\mu$m to parameterize these rates, as molecular simulation results for these are not available.  Figure S10 demonstrates that such source terms are necessary, since no anion dynamics appear after 4 hours of biasing due to the polarization of the anion vacancies without them. The sensitivity of the dynamics to the initial position of the composition front (Figs. S12 and S13 without and with vacancy source terms, respectively) does not ameliorate this, nor does changing the total amount of vacancies in the system (Figs. S14 and S15). We further find that the anion dynamics is only affected by the injection rate of vacancies at the cathode (Fig. S16), over the range of source terms considered that are simultaneously consistent with the assumption of linear dependence on applied voltage. 

In the reverse case, shown in Figure 2d, no significant change of the composition profile is observed, consistent with experiment. This is due to the suppressed vacancy source terms  when the cathode is applied to the Cl segment of the heterojunction. This lower rate is consistent with the higher vacancy formation energy of pure Cl lattices, which are 1.32 $e$V and 1.44 $e$V for pure Br and Cl lattices, respectively \cite{lai2018intrinsic}.  The simulations in Figure 2b and d are used for the numerical comparison in Figure 1e, where we see that for electric biases above 1.5 V/$\mu$m, the dynamics is dominated by the vacancy source terms at the electrodes' interfaces, while below -1.5 V/$\mu$m migration is prevented by vacancies being depleted in order to screen the field. Whereas, when the bias intensity is low, the migration length evolution becomes nonlinear due to the interplay of the interdiffusion process that tends to shift the front toward the bromide rich segment when no bias is applied\cite{lai2018intrinsic}, and the effects of the injection of vacancies at the cathode.  

\begin{figure}
\begin{center}
\includegraphics[width=8.5cm]{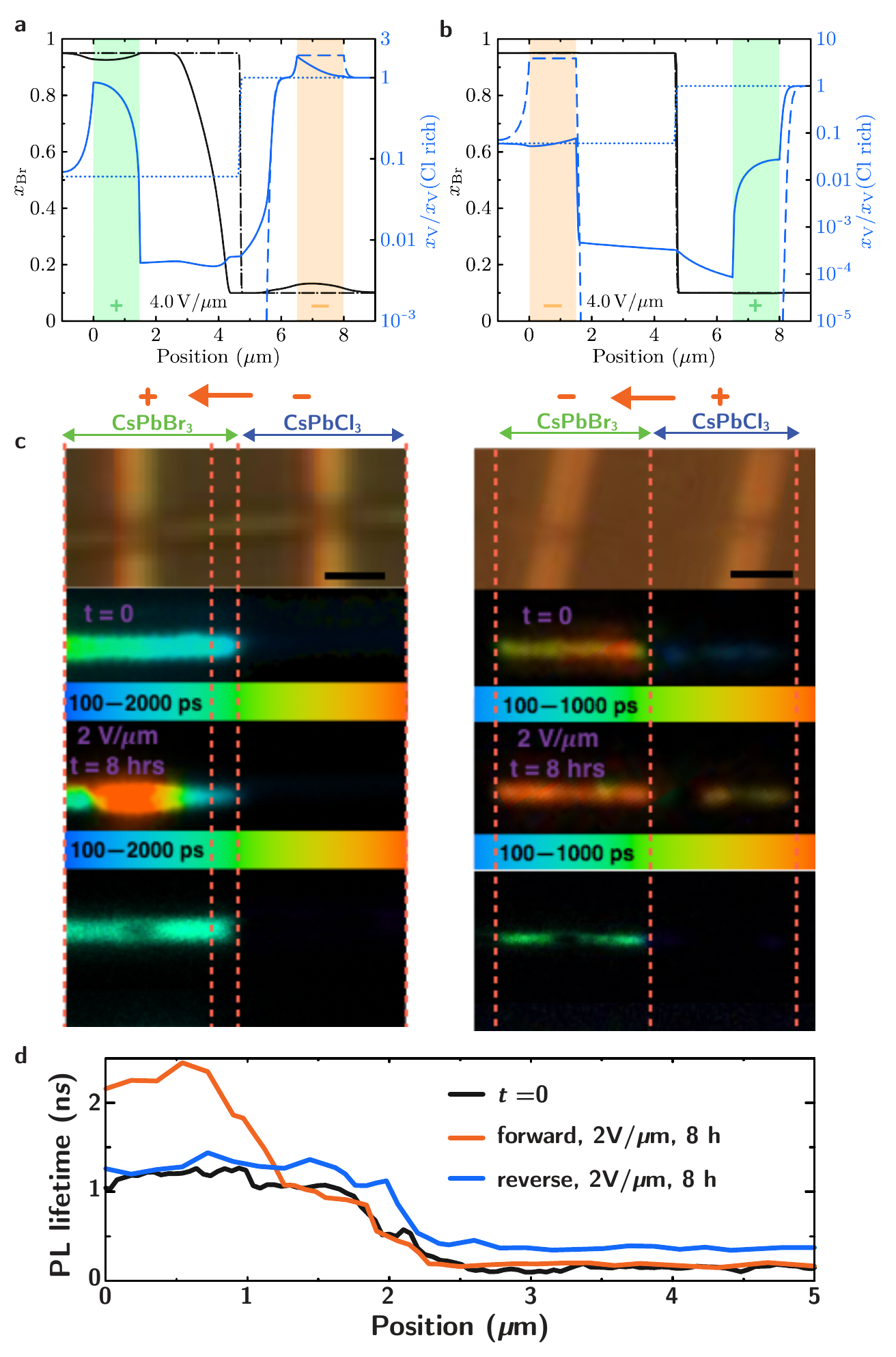}
\caption{Electric-field driven ion rectification mechanism modeling and analysis. Simulated halide composition (black) and vacancy concentration (blue) evolutions along the CsPbBr$_3$-CsPbCl$_3$ heterostructure nanowire under 4-hour biasing computed numerically (see text) with 4 V/$\mu$m bias intensity under forward (a) and reverse (b) conditions, respectively. The vacancy concentration is rescaled by its value for the chloride rich par of the wire. The dotted lines correspond to the initial profiles, the dashed lines to the profiles after 4- hours of biasing with no source terms at the electrodes' interfaces and the solid lines to the profiles after 4 hours of biasing with source terms ($\alpha=1.3 \times 10^{-5}  \mu$m V$^{-1}$ s$^{-1}$ in the forward case and $\alpha=3.8 \times 10^{-7} \mu$m V$^{-1}$ s$^{-1}$ in the reverse one that roughly correspond to the creation/destruction of 60 and 1.6 vacancy per 10$^3$ unit cells per hour with a bias of 1 V/$\mu$m in the forward and reverse biasing, respectively). The positions of the anode and the cathode are indicated by the orange and green zones represent the anode and cathode positions, respectively. The The width of the electrode widths is 1.5 $\mu$m.  (c) Representative fluorescence lifetime and confocal PL images of the heterostructure nanowire under different bias polarity. Increased lifetime corresponds indicates to the decreased of halide vacancy concentrations, indicating halide ions are accumulated under forward bias condition. While due to the interdiffusion under reverse bias, the lifetime on both segments slightly increases. The scale bar for all images is 2 $\mu$m. (d) Fluorescence lifetime profile along the heterostructure nanowire within the parallel Au electrodes.}
\label{Fi:3}
\end{center} 
\end{figure}

In Figures 3a and 3b, we show that the injection of vacancies at the cathode allows for nonzero vacancy concentration in the otherwise depleted zone, which allows anion interdiffusion/migration to occur. Additional biases are considered in the Supplementary Materials (see Fig. S11). The injection rate at the cathode mainly controls the vacancy concentration in this otherwise depleted zone. Increasing the injection rate increases the vacancy concentration, which allows for an increasing drift of the anion concentration front toward the bromide rich part of the wire in the forward bias case. Although, the details of the vacancy concentration profile in the otherwise depleted zone depends on the anion composition profile, as seen in Figure 3a around the composition front.  If the vacancy concentration close to the initial composition front stays sufficiently low despite the additional contribution due to the vacancy sources, no interdiffusion occurs as illustrated in Figure 3b. 

The hypothesis of vacancy-driven ion rectification under electric field is supported by fluorescence lifetime measurements along the heterostructure nanowire under different electric field direction. The confocal PL characterization was first conducted to confirm the electric-field effect on the heterostructure nanowire, as shown in Figure 3c, in which the interface shifted under the forward bias, while barely moved under reverse bias. The fluorescence lifetime is determined by the carrier recombination rate at a specific location, which is closely related to the vacancy concentration in the lattice due to trap assisted recombination pathways. As indicated by the different halide vacancy concentrations of the two segments, the lifetime on CsPbBr$_3$ segment is much longer than that on CsPbCl$_3$ one. Based on the assumption that longer fluorescence lifetime is due to lower vacancy concentration, we attributed the significant increase of lifetime on the pure CsPbBr$_3$ segment under forward bias to the migration of halide ions and the depletion of halide vacancies. While, when we switched the bias polarity, the fluorescence lifetime on both segments slightly increase due to the interdiffusion of halide ions. The direct comparison of fluorescence lifetime along the nanowire under different bias polarities is presented in Figure 3d, consistent with the perspective of vacancy-assisted ionic rectification in the nanowire heterostructure.

In summary, we have successfully visualized the halide ion migration under the electric field in well-defined CsPbBr$_3$-CsPbCl$_3$ nanowire heterostructures. Through the investigation of the halide ion migration dynamics across the heterointerface under applied voltage, we have quantified the halide ion mobility to be $1.7 \times10^{-13} \mathrm{cm}^2 V^{-1} \mathrm{s}^{-1}$ at the room temperature and revealed a unique ionic rectification phenomenon for the first time in a lead-halide based system.  We have shown that this rectification effect is independent of the sample preparation, and instead related to the injection of halide vacancies at the cathode. Theoretical simulation based on the vacancy-assisted ion migration mechanism exhibits consistent results with the experimental observation, suggesting that vacancy creation and destruction at the electrodes' interfaces is crucial for the field-induced ionic rectification, and proceeds due to the high ionic mobility and concomitant composition dependence. The demonstration of such ionic modulation not only shows the promise of using dynamic halide perovskite ionic lattice for ionic device applications, but also sheds light on the fundamental mechanistic understanding of the asymmetric ionic transport in solid-state systems generally. 

\vspace{0.5cm}

{\bf Supporting Materials:} Can be found \href{http://www.cchem.berkeley.edu/dtlgrp/DL51.pdf}{http://www.cchem.berkeley.edu/dtlgrp/DL51.pdf}.

\vspace{0.5cm}

{\bf Acknowledgments:}
The authors are grateful to Dr. J. Kang, Y. Zhang for fruitful discussions. We thank the nanofabrication and imaging facilities in Marvell Nanofabrication Laboratory and Molecular Foundry. Funding: This work was supported by the U. S. Department of Energy, Office of Science, Office of Basic Energy Sciences, Materials Sciences and Engineering Division, under Contract No. DE-AC02-05-CH11231 within the Physical Chemistry of Inorganic Nanostructures Program (KC3103). The FLIM experiments were conducted at the Cancer Reserch Laboratory Molecular Imaging Center, University of California, Berkeley, which was supported by NSF Grant DBI-0116016. The LSM 710 confocal microscopic experiments were conducted at the College of Natural Resources Biological Imaging Facility, University of California, Berkeley, which was supported in part by National Institutes of Health S10 Program under Award 1S10RR026866-01. The content is solely the responsibility of the authors and does not necessarily represent the official view of the National Institutes of Health. This research used resources of the National Energy Research Scientific Computing Center, a US Department of Energy Office of Science User Facility operated under Contract DE-AC02-05CH11231. Q.K and M.L. thank the fellowship support from Suzhou Industrial Park. 

{\bf Author contributions:} Q.K., A.O., M.L., D.T.L. and P.Y. conceived and designed the experiment. Q.K. fabricated heterojunction device and conducted electrical biasing experiment. M.L. synthesized nanowire samples and performed PL and FLIM measurements. M.G. developed the algorithm for automatic confocal PL data analysis. A.O. and D.T.L. developed theoretical model and conducted simulation. All authors discussed the results, analyzed data and wrote the manuscript. 

{\bf Competing interests:} The authors declare no competing interests. 

{\bf Data and materials availability:} All data is available in the main text or the supplementary materials.

%\bibliography{bib}

\begin{thebibliography}{33}%
\makeatletter
\providecommand \@ifxundefined [1]{%
 \@ifx{#1\undefined}
}%
\providecommand \@ifnum [1]{%
 \ifnum #1\expandafter \@firstoftwo
 \else \expandafter \@secondoftwo
 \fi
}%
\providecommand \@ifx [1]{%
 \ifx #1\expandafter \@firstoftwo
 \else \expandafter \@secondoftwo
 \fi
}%
\providecommand \natexlab [1]{#1}%
\providecommand \enquote  [1]{``#1''}%
\providecommand \bibnamefont  [1]{#1}%
\providecommand \bibfnamefont [1]{#1}%
\providecommand \citenamefont [1]{#1}%
\providecommand \href@noop [0]{\@secondoftwo}%
\providecommand \href [0]{\begingroup \@sanitize@url \@href}%
\providecommand \@href[1]{\@@startlink{#1}\@@href}%
\providecommand \@@href[1]{\endgroup#1\@@endlink}%
\providecommand \@sanitize@url [0]{\catcode `\\12\catcode `\$12\catcode
  `\&12\catcode `\#12\catcode `\^12\catcode `\_12\catcode `\%12\relax}%
\providecommand \@@startlink[1]{}%
\providecommand \@@endlink[0]{}%
\providecommand \url  [0]{\begingroup\@sanitize@url \@url }%
\providecommand \@url [1]{\endgroup\@href {#1}{\urlprefix }}%
\providecommand \urlprefix  [0]{URL }%
\providecommand \Eprint [0]{\href }%
\providecommand \doibase [0]{http://dx.doi.org/}%
\providecommand \selectlanguage [0]{\@gobble}%
\providecommand \bibinfo  [0]{\@secondoftwo}%
\providecommand \bibfield  [0]{\@secondoftwo}%
\providecommand \translation [1]{[#1]}%
\providecommand \BibitemOpen [0]{}%
\providecommand \bibitemStop [0]{}%
\providecommand \bibitemNoStop [0]{.\EOS\space}%
\providecommand \EOS [0]{\spacefactor3000\relax}%
\providecommand \BibitemShut  [1]{\csname bibitem#1\endcsname}%
\let\auto@bib@innerbib\@empty
%</preamble>
\bibitem [{\citenamefont {Brittman}\ \emph {et~al.}(2015)\citenamefont
  {Brittman}, \citenamefont {Adhyaksa},\ and\ \citenamefont
  {Garnett}}]{brittman2015expanding}%
  \BibitemOpen
  \bibfield  {author} {\bibinfo {author} {\bibfnamefont {S.}~\bibnamefont
  {Brittman}}, \bibinfo {author} {\bibfnamefont {G.~W.~P.}\ \bibnamefont
  {Adhyaksa}}, \ and\ \bibinfo {author} {\bibfnamefont {E.~C.}\ \bibnamefont
  {Garnett}},\ }\href@noop {} {\bibfield  {journal} {\bibinfo  {journal} {MRS
  communications}\ }\textbf {\bibinfo {volume} {5}},\ \bibinfo {pages} {7}
  (\bibinfo {year} {2015})}\BibitemShut {NoStop}%
\bibitem [{\citenamefont {Kojima}\ \emph {et~al.}(2009)\citenamefont {Kojima},
  \citenamefont {Teshima}, \citenamefont {Shirai},\ and\ \citenamefont
  {Miyasaka}}]{kojima2009organometal}%
  \BibitemOpen
  \bibfield  {author} {\bibinfo {author} {\bibfnamefont {A.}~\bibnamefont
  {Kojima}}, \bibinfo {author} {\bibfnamefont {K.}~\bibnamefont {Teshima}},
  \bibinfo {author} {\bibfnamefont {Y.}~\bibnamefont {Shirai}}, \ and\ \bibinfo
  {author} {\bibfnamefont {T.}~\bibnamefont {Miyasaka}},\ }\href@noop {}
  {\bibfield  {journal} {\bibinfo  {journal} {Journal of the American Chemical
  Society}\ }\textbf {\bibinfo {volume} {131}},\ \bibinfo {pages} {6050}
  (\bibinfo {year} {2009})}\BibitemShut {NoStop}%
\bibitem [{\citenamefont {Green}\ \emph {et~al.}(2014)\citenamefont {Green},
  \citenamefont {Ho-Baillie},\ and\ \citenamefont
  {Snaith}}]{green2014emergence}%
  \BibitemOpen
  \bibfield  {author} {\bibinfo {author} {\bibfnamefont {M.~A.}\ \bibnamefont
  {Green}}, \bibinfo {author} {\bibfnamefont {A.}~\bibnamefont {Ho-Baillie}}, \
  and\ \bibinfo {author} {\bibfnamefont {H.~J.}\ \bibnamefont {Snaith}},\
  }\href@noop {} {\bibfield  {journal} {\bibinfo  {journal} {Nature photonics}\
  }\textbf {\bibinfo {volume} {8}},\ \bibinfo {pages} {506} (\bibinfo {year}
  {2014})}\BibitemShut {NoStop}%
\bibitem [{\citenamefont {Lee}\ \emph {et~al.}(2012)\citenamefont {Lee},
  \citenamefont {Teuscher}, \citenamefont {Miyasaka}, \citenamefont
  {Murakami},\ and\ \citenamefont {Snaith}}]{lee2012efficient}%
  \BibitemOpen
  \bibfield  {author} {\bibinfo {author} {\bibfnamefont {M.~M.}\ \bibnamefont
  {Lee}}, \bibinfo {author} {\bibfnamefont {J.}~\bibnamefont {Teuscher}},
  \bibinfo {author} {\bibfnamefont {T.}~\bibnamefont {Miyasaka}}, \bibinfo
  {author} {\bibfnamefont {T.~N.}\ \bibnamefont {Murakami}}, \ and\ \bibinfo
  {author} {\bibfnamefont {H.~J.}\ \bibnamefont {Snaith}},\ }\href@noop {}
  {\bibfield  {journal} {\bibinfo  {journal} {Science}\ }\textbf {\bibinfo
  {volume} {338}},\ \bibinfo {pages} {643} (\bibinfo {year}
  {2012})}\BibitemShut {NoStop}%
\bibitem [{\citenamefont {Xing}\ \emph {et~al.}(2014)\citenamefont {Xing},
  \citenamefont {Mathews}, \citenamefont {Lim}, \citenamefont {Yantara},
  \citenamefont {Liu}, \citenamefont {Sabba}, \citenamefont {Gr{\"a}tzel},
  \citenamefont {Mhaisalkar},\ and\ \citenamefont {Sum}}]{xing2014low}%
  \BibitemOpen
  \bibfield  {author} {\bibinfo {author} {\bibfnamefont {G.}~\bibnamefont
  {Xing}}, \bibinfo {author} {\bibfnamefont {N.}~\bibnamefont {Mathews}},
  \bibinfo {author} {\bibfnamefont {S.~S.}\ \bibnamefont {Lim}}, \bibinfo
  {author} {\bibfnamefont {N.}~\bibnamefont {Yantara}}, \bibinfo {author}
  {\bibfnamefont {X.}~\bibnamefont {Liu}}, \bibinfo {author} {\bibfnamefont
  {D.}~\bibnamefont {Sabba}}, \bibinfo {author} {\bibfnamefont
  {M.}~\bibnamefont {Gr{\"a}tzel}}, \bibinfo {author} {\bibfnamefont
  {S.}~\bibnamefont {Mhaisalkar}}, \ and\ \bibinfo {author} {\bibfnamefont
  {T.~C.}\ \bibnamefont {Sum}},\ }\href@noop {} {\bibfield  {journal} {\bibinfo
   {journal} {Nature materials}\ }\textbf {\bibinfo {volume} {13}},\ \bibinfo
  {pages} {476} (\bibinfo {year} {2014})}\BibitemShut {NoStop}%
\bibitem [{\citenamefont {Zhu}\ \emph {et~al.}(2015)\citenamefont {Zhu},
  \citenamefont {Fu}, \citenamefont {Meng}, \citenamefont {Wu}, \citenamefont
  {Gong}, \citenamefont {Ding}, \citenamefont {Gustafsson}, \citenamefont
  {Trinh}, \citenamefont {Jin},\ and\ \citenamefont {Zhu}}]{zhu2015lead}%
  \BibitemOpen
  \bibfield  {author} {\bibinfo {author} {\bibfnamefont {H.}~\bibnamefont
  {Zhu}}, \bibinfo {author} {\bibfnamefont {Y.}~\bibnamefont {Fu}}, \bibinfo
  {author} {\bibfnamefont {F.}~\bibnamefont {Meng}}, \bibinfo {author}
  {\bibfnamefont {X.}~\bibnamefont {Wu}}, \bibinfo {author} {\bibfnamefont
  {Z.}~\bibnamefont {Gong}}, \bibinfo {author} {\bibfnamefont {Q.}~\bibnamefont
  {Ding}}, \bibinfo {author} {\bibfnamefont {M.~V.}\ \bibnamefont
  {Gustafsson}}, \bibinfo {author} {\bibfnamefont {M.~T.}\ \bibnamefont
  {Trinh}}, \bibinfo {author} {\bibfnamefont {S.}~\bibnamefont {Jin}}, \ and\
  \bibinfo {author} {\bibfnamefont {X.}~\bibnamefont {Zhu}},\ }\href@noop {}
  {\bibfield  {journal} {\bibinfo  {journal} {Nature materials}\ }\textbf
  {\bibinfo {volume} {14}},\ \bibinfo {pages} {636} (\bibinfo {year}
  {2015})}\BibitemShut {NoStop}%
\bibitem [{\citenamefont {Cho}\ \emph {et~al.}(2015)\citenamefont {Cho},
  \citenamefont {Jeong}, \citenamefont {Park}, \citenamefont {Kim},
  \citenamefont {Wolf}, \citenamefont {Lee}, \citenamefont {Heo}, \citenamefont
  {Sadhanala}, \citenamefont {Myoung}, \citenamefont {Yoo} \emph
  {et~al.}}]{cho2015overcoming}%
  \BibitemOpen
  \bibfield  {author} {\bibinfo {author} {\bibfnamefont {H.}~\bibnamefont
  {Cho}}, \bibinfo {author} {\bibfnamefont {S.-H.}\ \bibnamefont {Jeong}},
  \bibinfo {author} {\bibfnamefont {M.-H.}\ \bibnamefont {Park}}, \bibinfo
  {author} {\bibfnamefont {Y.-H.}\ \bibnamefont {Kim}}, \bibinfo {author}
  {\bibfnamefont {C.}~\bibnamefont {Wolf}}, \bibinfo {author} {\bibfnamefont
  {C.-L.}\ \bibnamefont {Lee}}, \bibinfo {author} {\bibfnamefont {J.~H.}\
  \bibnamefont {Heo}}, \bibinfo {author} {\bibfnamefont {A.}~\bibnamefont
  {Sadhanala}}, \bibinfo {author} {\bibfnamefont {N.}~\bibnamefont {Myoung}},
  \bibinfo {author} {\bibfnamefont {S.}~\bibnamefont {Yoo}},  \emph {et~al.},\
  }\href@noop {} {\bibfield  {journal} {\bibinfo  {journal} {Science}\ }\textbf
  {\bibinfo {volume} {350}},\ \bibinfo {pages} {1222} (\bibinfo {year}
  {2015})}\BibitemShut {NoStop}%
\bibitem [{\citenamefont {Lin}\ \emph {et~al.}(2018{\natexlab{a}})\citenamefont
  {Lin}, \citenamefont {Xing}, \citenamefont {Quan}, \citenamefont {de~Arquer},
  \citenamefont {Gong}, \citenamefont {Lu}, \citenamefont {Xie}, \citenamefont
  {Zhao}, \citenamefont {Zhang}, \citenamefont {Yan} \emph
  {et~al.}}]{lin2018perovskite}%
  \BibitemOpen
  \bibfield  {author} {\bibinfo {author} {\bibfnamefont {K.}~\bibnamefont
  {Lin}}, \bibinfo {author} {\bibfnamefont {J.}~\bibnamefont {Xing}}, \bibinfo
  {author} {\bibfnamefont {L.~N.}\ \bibnamefont {Quan}}, \bibinfo {author}
  {\bibfnamefont {F.~P.~G.}\ \bibnamefont {de~Arquer}}, \bibinfo {author}
  {\bibfnamefont {X.}~\bibnamefont {Gong}}, \bibinfo {author} {\bibfnamefont
  {J.}~\bibnamefont {Lu}}, \bibinfo {author} {\bibfnamefont {L.}~\bibnamefont
  {Xie}}, \bibinfo {author} {\bibfnamefont {W.}~\bibnamefont {Zhao}}, \bibinfo
  {author} {\bibfnamefont {D.}~\bibnamefont {Zhang}}, \bibinfo {author}
  {\bibfnamefont {C.}~\bibnamefont {Yan}},  \emph {et~al.},\ }\href@noop {}
  {\bibfield  {journal} {\bibinfo  {journal} {Nature}\ }\textbf {\bibinfo
  {volume} {562}},\ \bibinfo {pages} {245} (\bibinfo {year}
  {2018}{\natexlab{a}})}\BibitemShut {NoStop}%
\bibitem [{\citenamefont {Cao}\ \emph {et~al.}(2018)\citenamefont {Cao},
  \citenamefont {Wang}, \citenamefont {Tian}, \citenamefont {Guo},
  \citenamefont {Wei}, \citenamefont {Chen}, \citenamefont {Miao},
  \citenamefont {Zou}, \citenamefont {Pan}, \citenamefont {He} \emph
  {et~al.}}]{cao2018perovskite}%
  \BibitemOpen
  \bibfield  {author} {\bibinfo {author} {\bibfnamefont {Y.}~\bibnamefont
  {Cao}}, \bibinfo {author} {\bibfnamefont {N.}~\bibnamefont {Wang}}, \bibinfo
  {author} {\bibfnamefont {H.}~\bibnamefont {Tian}}, \bibinfo {author}
  {\bibfnamefont {J.}~\bibnamefont {Guo}}, \bibinfo {author} {\bibfnamefont
  {Y.}~\bibnamefont {Wei}}, \bibinfo {author} {\bibfnamefont {H.}~\bibnamefont
  {Chen}}, \bibinfo {author} {\bibfnamefont {Y.}~\bibnamefont {Miao}}, \bibinfo
  {author} {\bibfnamefont {W.}~\bibnamefont {Zou}}, \bibinfo {author}
  {\bibfnamefont {K.}~\bibnamefont {Pan}}, \bibinfo {author} {\bibfnamefont
  {Y.}~\bibnamefont {He}},  \emph {et~al.},\ }\href@noop {} {\bibfield
  {journal} {\bibinfo  {journal} {Nature}\ }\textbf {\bibinfo {volume} {562}},\
  \bibinfo {pages} {249} (\bibinfo {year} {2018})}\BibitemShut {NoStop}%
\bibitem [{\citenamefont {Wang}\ \emph {et~al.}(2016)\citenamefont {Wang},
  \citenamefont {Cheng}, \citenamefont {Ge}, \citenamefont {Zhang},
  \citenamefont {Miao}, \citenamefont {Zou}, \citenamefont {Yi}, \citenamefont
  {Sun}, \citenamefont {Cao}, \citenamefont {Yang} \emph
  {et~al.}}]{wang2016perovskite}%
  \BibitemOpen
  \bibfield  {author} {\bibinfo {author} {\bibfnamefont {N.}~\bibnamefont
  {Wang}}, \bibinfo {author} {\bibfnamefont {L.}~\bibnamefont {Cheng}},
  \bibinfo {author} {\bibfnamefont {R.}~\bibnamefont {Ge}}, \bibinfo {author}
  {\bibfnamefont {S.}~\bibnamefont {Zhang}}, \bibinfo {author} {\bibfnamefont
  {Y.}~\bibnamefont {Miao}}, \bibinfo {author} {\bibfnamefont {W.}~\bibnamefont
  {Zou}}, \bibinfo {author} {\bibfnamefont {C.}~\bibnamefont {Yi}}, \bibinfo
  {author} {\bibfnamefont {Y.}~\bibnamefont {Sun}}, \bibinfo {author}
  {\bibfnamefont {Y.}~\bibnamefont {Cao}}, \bibinfo {author} {\bibfnamefont
  {R.}~\bibnamefont {Yang}},  \emph {et~al.},\ }\href@noop {} {\bibfield
  {journal} {\bibinfo  {journal} {Nature Photonics}\ }\textbf {\bibinfo
  {volume} {10}},\ \bibinfo {pages} {699} (\bibinfo {year} {2016})}\BibitemShut
  {NoStop}%
\bibitem [{\citenamefont {Dou}\ \emph {et~al.}(2017)\citenamefont {Dou},
  \citenamefont {Lai}, \citenamefont {Kley}, \citenamefont {Yang},
  \citenamefont {Bischak}, \citenamefont {Zhang}, \citenamefont {Eaton},
  \citenamefont {Ginsberg},\ and\ \citenamefont {Yang}}]{dou2017spatially}%
  \BibitemOpen
  \bibfield  {author} {\bibinfo {author} {\bibfnamefont {L.}~\bibnamefont
  {Dou}}, \bibinfo {author} {\bibfnamefont {M.}~\bibnamefont {Lai}}, \bibinfo
  {author} {\bibfnamefont {C.~S.}\ \bibnamefont {Kley}}, \bibinfo {author}
  {\bibfnamefont {Y.}~\bibnamefont {Yang}}, \bibinfo {author} {\bibfnamefont
  {C.~G.}\ \bibnamefont {Bischak}}, \bibinfo {author} {\bibfnamefont
  {D.}~\bibnamefont {Zhang}}, \bibinfo {author} {\bibfnamefont {S.~W.}\
  \bibnamefont {Eaton}}, \bibinfo {author} {\bibfnamefont {N.~S.}\ \bibnamefont
  {Ginsberg}}, \ and\ \bibinfo {author} {\bibfnamefont {P.}~\bibnamefont
  {Yang}},\ }\href@noop {} {\bibfield  {journal} {\bibinfo  {journal}
  {Proceedings of the National Academy of Sciences}\ }\textbf {\bibinfo
  {volume} {114}},\ \bibinfo {pages} {7216} (\bibinfo {year}
  {2017})}\BibitemShut {NoStop}%
\bibitem [{\citenamefont {Wang}\ \emph {et~al.}(2017)\citenamefont {Wang},
  \citenamefont {Chen}, \citenamefont {Deschler}, \citenamefont {Sun},
  \citenamefont {Lu}, \citenamefont {Wertz}, \citenamefont {Hu},\ and\
  \citenamefont {Shi}}]{wang2017epitaxial}%
  \BibitemOpen
  \bibfield  {author} {\bibinfo {author} {\bibfnamefont {Y.}~\bibnamefont
  {Wang}}, \bibinfo {author} {\bibfnamefont {Z.}~\bibnamefont {Chen}}, \bibinfo
  {author} {\bibfnamefont {F.}~\bibnamefont {Deschler}}, \bibinfo {author}
  {\bibfnamefont {X.}~\bibnamefont {Sun}}, \bibinfo {author} {\bibfnamefont
  {T.-M.}\ \bibnamefont {Lu}}, \bibinfo {author} {\bibfnamefont {E.~A.}\
  \bibnamefont {Wertz}}, \bibinfo {author} {\bibfnamefont {J.-M.}\ \bibnamefont
  {Hu}}, \ and\ \bibinfo {author} {\bibfnamefont {J.}~\bibnamefont {Shi}},\
  }\href@noop {} {\bibfield  {journal} {\bibinfo  {journal} {ACS nano}\
  }\textbf {\bibinfo {volume} {11}},\ \bibinfo {pages} {3355} (\bibinfo {year}
  {2017})}\BibitemShut {NoStop}%
\bibitem [{\citenamefont {Kong}\ \emph {et~al.}(2018)\citenamefont {Kong},
  \citenamefont {Lee}, \citenamefont {Lai}, \citenamefont {Bischak},
  \citenamefont {Gao}, \citenamefont {Wong}, \citenamefont {Lei}, \citenamefont
  {Yu}, \citenamefont {Wang}, \citenamefont {Ginsberg} \emph
  {et~al.}}]{kong2018phase}%
  \BibitemOpen
  \bibfield  {author} {\bibinfo {author} {\bibfnamefont {Q.}~\bibnamefont
  {Kong}}, \bibinfo {author} {\bibfnamefont {W.}~\bibnamefont {Lee}}, \bibinfo
  {author} {\bibfnamefont {M.}~\bibnamefont {Lai}}, \bibinfo {author}
  {\bibfnamefont {C.~G.}\ \bibnamefont {Bischak}}, \bibinfo {author}
  {\bibfnamefont {G.}~\bibnamefont {Gao}}, \bibinfo {author} {\bibfnamefont
  {A.~B.}\ \bibnamefont {Wong}}, \bibinfo {author} {\bibfnamefont
  {T.}~\bibnamefont {Lei}}, \bibinfo {author} {\bibfnamefont {Y.}~\bibnamefont
  {Yu}}, \bibinfo {author} {\bibfnamefont {L.-W.}\ \bibnamefont {Wang}},
  \bibinfo {author} {\bibfnamefont {N.~S.}\ \bibnamefont {Ginsberg}},  \emph
  {et~al.},\ }\href@noop {} {\bibfield  {journal} {\bibinfo  {journal}
  {Proceedings of the National Academy of Sciences}\ }\textbf {\bibinfo
  {volume} {115}},\ \bibinfo {pages} {8889} (\bibinfo {year}
  {2018})}\BibitemShut {NoStop}%
\bibitem [{\citenamefont {Pan}\ \emph {et~al.}(2018)\citenamefont {Pan},
  \citenamefont {Fu}, \citenamefont {Chen}, \citenamefont {Czech},
  \citenamefont {Wright},\ and\ \citenamefont {Jin}}]{pan2018visualization}%
  \BibitemOpen
  \bibfield  {author} {\bibinfo {author} {\bibfnamefont {D.}~\bibnamefont
  {Pan}}, \bibinfo {author} {\bibfnamefont {Y.}~\bibnamefont {Fu}}, \bibinfo
  {author} {\bibfnamefont {J.}~\bibnamefont {Chen}}, \bibinfo {author}
  {\bibfnamefont {K.~J.}\ \bibnamefont {Czech}}, \bibinfo {author}
  {\bibfnamefont {J.~C.}\ \bibnamefont {Wright}}, \ and\ \bibinfo {author}
  {\bibfnamefont {S.}~\bibnamefont {Jin}},\ }\href@noop {} {\bibfield
  {journal} {\bibinfo  {journal} {Nano letters}\ }\textbf {\bibinfo {volume}
  {18}},\ \bibinfo {pages} {1807} (\bibinfo {year} {2018})}\BibitemShut
  {NoStop}%
\bibitem [{\citenamefont {Nedelcu}\ \emph {et~al.}(2015)\citenamefont
  {Nedelcu}, \citenamefont {Protesescu}, \citenamefont {Yakunin}, \citenamefont
  {Bodnarchuk}, \citenamefont {Grotevent},\ and\ \citenamefont
  {Kovalenko}}]{nedelcu2015fast}%
  \BibitemOpen
  \bibfield  {author} {\bibinfo {author} {\bibfnamefont {G.}~\bibnamefont
  {Nedelcu}}, \bibinfo {author} {\bibfnamefont {L.}~\bibnamefont {Protesescu}},
  \bibinfo {author} {\bibfnamefont {S.}~\bibnamefont {Yakunin}}, \bibinfo
  {author} {\bibfnamefont {M.~I.}\ \bibnamefont {Bodnarchuk}}, \bibinfo
  {author} {\bibfnamefont {M.~J.}\ \bibnamefont {Grotevent}}, \ and\ \bibinfo
  {author} {\bibfnamefont {M.~V.}\ \bibnamefont {Kovalenko}},\ }\href@noop {}
  {\bibfield  {journal} {\bibinfo  {journal} {Nano letters}\ }\textbf {\bibinfo
  {volume} {15}},\ \bibinfo {pages} {5635} (\bibinfo {year}
  {2015})}\BibitemShut {NoStop}%
\bibitem [{\citenamefont {Tress}(2017)}]{tress2017metal}%
  \BibitemOpen
  \bibfield  {author} {\bibinfo {author} {\bibfnamefont {W.}~\bibnamefont
  {Tress}},\ }\href@noop {} {\bibfield  {journal} {\bibinfo  {journal} {The
  journal of physical chemistry letters}\ }\textbf {\bibinfo {volume} {8}},\
  \bibinfo {pages} {3106} (\bibinfo {year} {2017})}\BibitemShut {NoStop}%
\bibitem [{\citenamefont {Mizusaki}\ \emph {et~al.}(1983)\citenamefont
  {Mizusaki}, \citenamefont {Arai},\ and\ \citenamefont
  {Fueki}}]{mizusaki1983ionic}%
  \BibitemOpen
  \bibfield  {author} {\bibinfo {author} {\bibfnamefont {J.}~\bibnamefont
  {Mizusaki}}, \bibinfo {author} {\bibfnamefont {K.}~\bibnamefont {Arai}}, \
  and\ \bibinfo {author} {\bibfnamefont {K.}~\bibnamefont {Fueki}},\
  }\href@noop {} {\bibfield  {journal} {\bibinfo  {journal} {Solid State
  Ionics}\ }\textbf {\bibinfo {volume} {11}},\ \bibinfo {pages} {203} (\bibinfo
  {year} {1983})}\BibitemShut {NoStop}%
\bibitem [{\citenamefont {Xiao}\ \emph {et~al.}(2015)\citenamefont {Xiao},
  \citenamefont {Yuan}, \citenamefont {Shao}, \citenamefont {Wang},
  \citenamefont {Dong}, \citenamefont {Bi}, \citenamefont {Sharma},
  \citenamefont {Gruverman},\ and\ \citenamefont {Huang}}]{xiao2015giant}%
  \BibitemOpen
  \bibfield  {author} {\bibinfo {author} {\bibfnamefont {Z.}~\bibnamefont
  {Xiao}}, \bibinfo {author} {\bibfnamefont {Y.}~\bibnamefont {Yuan}}, \bibinfo
  {author} {\bibfnamefont {Y.}~\bibnamefont {Shao}}, \bibinfo {author}
  {\bibfnamefont {Q.}~\bibnamefont {Wang}}, \bibinfo {author} {\bibfnamefont
  {Q.}~\bibnamefont {Dong}}, \bibinfo {author} {\bibfnamefont {C.}~\bibnamefont
  {Bi}}, \bibinfo {author} {\bibfnamefont {P.}~\bibnamefont {Sharma}}, \bibinfo
  {author} {\bibfnamefont {A.}~\bibnamefont {Gruverman}}, \ and\ \bibinfo
  {author} {\bibfnamefont {J.}~\bibnamefont {Huang}},\ }\href@noop {}
  {\bibfield  {journal} {\bibinfo  {journal} {Nature materials}\ }\textbf
  {\bibinfo {volume} {14}},\ \bibinfo {pages} {193} (\bibinfo {year}
  {2015})}\BibitemShut {NoStop}%
\bibitem [{\citenamefont {Game}\ \emph {et~al.}(2017)\citenamefont {Game},
  \citenamefont {Buchsbaum}, \citenamefont {Zhou}, \citenamefont {Padture},\
  and\ \citenamefont {Kingon}}]{game2017ions}%
  \BibitemOpen
  \bibfield  {author} {\bibinfo {author} {\bibfnamefont {O.~S.}\ \bibnamefont
  {Game}}, \bibinfo {author} {\bibfnamefont {G.~J.}\ \bibnamefont {Buchsbaum}},
  \bibinfo {author} {\bibfnamefont {Y.}~\bibnamefont {Zhou}}, \bibinfo {author}
  {\bibfnamefont {N.~P.}\ \bibnamefont {Padture}}, \ and\ \bibinfo {author}
  {\bibfnamefont {A.~I.}\ \bibnamefont {Kingon}},\ }\href@noop {} {\bibfield
  {journal} {\bibinfo  {journal} {Advanced Functional Materials}\ }\textbf
  {\bibinfo {volume} {27}},\ \bibinfo {pages} {1606584} (\bibinfo {year}
  {2017})}\BibitemShut {NoStop}%
\bibitem [{\citenamefont {Calado}\ \emph {et~al.}(2016)\citenamefont {Calado},
  \citenamefont {Telford}, \citenamefont {Bryant}, \citenamefont {Li},
  \citenamefont {Nelson}, \citenamefont {O’Regan},\ and\ \citenamefont
  {Barnes}}]{calado2016evidence}%
  \BibitemOpen
  \bibfield  {author} {\bibinfo {author} {\bibfnamefont {P.}~\bibnamefont
  {Calado}}, \bibinfo {author} {\bibfnamefont {A.~M.}\ \bibnamefont {Telford}},
  \bibinfo {author} {\bibfnamefont {D.}~\bibnamefont {Bryant}}, \bibinfo
  {author} {\bibfnamefont {X.}~\bibnamefont {Li}}, \bibinfo {author}
  {\bibfnamefont {J.}~\bibnamefont {Nelson}}, \bibinfo {author} {\bibfnamefont
  {B.~C.}\ \bibnamefont {O’Regan}}, \ and\ \bibinfo {author} {\bibfnamefont
  {P.~R.}\ \bibnamefont {Barnes}},\ }\href@noop {} {\bibfield  {journal}
  {\bibinfo  {journal} {Nature communications}\ }\textbf {\bibinfo {volume}
  {7}},\ \bibinfo {pages} {1} (\bibinfo {year} {2016})}\BibitemShut {NoStop}%
\bibitem [{\citenamefont {Lai}\ \emph {et~al.}(2018)\citenamefont {Lai},
  \citenamefont {Obliger}, \citenamefont {Lu}, \citenamefont {Kley},
  \citenamefont {Bischak}, \citenamefont {Kong}, \citenamefont {Lei},
  \citenamefont {Dou}, \citenamefont {Ginsberg}, \citenamefont {Limmer} \emph
  {et~al.}}]{lai2018intrinsic}%
  \BibitemOpen
  \bibfield  {author} {\bibinfo {author} {\bibfnamefont {M.}~\bibnamefont
  {Lai}}, \bibinfo {author} {\bibfnamefont {A.}~\bibnamefont {Obliger}},
  \bibinfo {author} {\bibfnamefont {D.}~\bibnamefont {Lu}}, \bibinfo {author}
  {\bibfnamefont {C.~S.}\ \bibnamefont {Kley}}, \bibinfo {author}
  {\bibfnamefont {C.~G.}\ \bibnamefont {Bischak}}, \bibinfo {author}
  {\bibfnamefont {Q.}~\bibnamefont {Kong}}, \bibinfo {author} {\bibfnamefont
  {T.}~\bibnamefont {Lei}}, \bibinfo {author} {\bibfnamefont {L.}~\bibnamefont
  {Dou}}, \bibinfo {author} {\bibfnamefont {N.~S.}\ \bibnamefont {Ginsberg}},
  \bibinfo {author} {\bibfnamefont {D.~T.}\ \bibnamefont {Limmer}},  \emph
  {et~al.},\ }\href@noop {} {\bibfield  {journal} {\bibinfo  {journal}
  {Proceedings of the National Academy of Sciences}\ }\textbf {\bibinfo
  {volume} {115}},\ \bibinfo {pages} {11929} (\bibinfo {year}
  {2018})}\BibitemShut {NoStop}%
\bibitem [{\citenamefont {Lin}\ \emph {et~al.}(2018{\natexlab{b}})\citenamefont
  {Lin}, \citenamefont {Yeh},\ and\ \citenamefont {Siwy}}]{lin2018voltage}%
  \BibitemOpen
  \bibfield  {author} {\bibinfo {author} {\bibfnamefont {C.-Y.}\ \bibnamefont
  {Lin}}, \bibinfo {author} {\bibfnamefont {L.-H.}\ \bibnamefont {Yeh}}, \ and\
  \bibinfo {author} {\bibfnamefont {Z.~S.}\ \bibnamefont {Siwy}},\ }\href@noop
  {} {\bibfield  {journal} {\bibinfo  {journal} {The journal of physical
  chemistry letters}\ }\textbf {\bibinfo {volume} {9}},\ \bibinfo {pages} {393}
  (\bibinfo {year} {2018}{\natexlab{b}})}\BibitemShut {NoStop}%
\bibitem [{\citenamefont {Cheng}\ and\ \citenamefont
  {Guo}(2009)}]{cheng2009ionic}%
  \BibitemOpen
  \bibfield  {author} {\bibinfo {author} {\bibfnamefont {L.-J.}\ \bibnamefont
  {Cheng}}\ and\ \bibinfo {author} {\bibfnamefont {L.~J.}\ \bibnamefont
  {Guo}},\ }\href@noop {} {\bibfield  {journal} {\bibinfo  {journal} {ACS
  nano}\ }\textbf {\bibinfo {volume} {3}},\ \bibinfo {pages} {575} (\bibinfo
  {year} {2009})}\BibitemShut {NoStop}%
\bibitem [{\citenamefont {Yan}\ \emph {et~al.}(2009)\citenamefont {Yan},
  \citenamefont {Liang}, \citenamefont {Fan},\ and\ \citenamefont
  {Yang}}]{yan2009nanofluidic}%
  \BibitemOpen
  \bibfield  {author} {\bibinfo {author} {\bibfnamefont {R.}~\bibnamefont
  {Yan}}, \bibinfo {author} {\bibfnamefont {W.}~\bibnamefont {Liang}}, \bibinfo
  {author} {\bibfnamefont {R.}~\bibnamefont {Fan}}, \ and\ \bibinfo {author}
  {\bibfnamefont {P.}~\bibnamefont {Yang}},\ }\href@noop {} {\bibfield
  {journal} {\bibinfo  {journal} {Nano letters}\ }\textbf {\bibinfo {volume}
  {9}},\ \bibinfo {pages} {3820} (\bibinfo {year} {2009})}\BibitemShut
  {NoStop}%
\bibitem [{\citenamefont {Fan}\ \emph {et~al.}(2005)\citenamefont {Fan},
  \citenamefont {Yue}, \citenamefont {Karnik}, \citenamefont {Majumdar},\ and\
  \citenamefont {Yang}}]{fan2005polarity}%
  \BibitemOpen
  \bibfield  {author} {\bibinfo {author} {\bibfnamefont {R.}~\bibnamefont
  {Fan}}, \bibinfo {author} {\bibfnamefont {M.}~\bibnamefont {Yue}}, \bibinfo
  {author} {\bibfnamefont {R.}~\bibnamefont {Karnik}}, \bibinfo {author}
  {\bibfnamefont {A.}~\bibnamefont {Majumdar}}, \ and\ \bibinfo {author}
  {\bibfnamefont {P.}~\bibnamefont {Yang}},\ }\href@noop {} {\bibfield
  {journal} {\bibinfo  {journal} {Physical review letters}\ }\textbf {\bibinfo
  {volume} {95}},\ \bibinfo {pages} {086607} (\bibinfo {year}
  {2005})}\BibitemShut {NoStop}%
\bibitem [{\citenamefont {Zhang}\ \emph {et~al.}(2016)\citenamefont {Zhang},
  \citenamefont {Wang}, \citenamefont {Xu}, \citenamefont {Liu}, \citenamefont
  {Song}, \citenamefont {Xue}, \citenamefont {Wang}, \citenamefont {Zheng},
  \citenamefont {Jiang}, \citenamefont {Zheng} \emph
  {et~al.}}]{zhang2016reversible}%
  \BibitemOpen
  \bibfield  {author} {\bibinfo {author} {\bibfnamefont {Y.}~\bibnamefont
  {Zhang}}, \bibinfo {author} {\bibfnamefont {Y.}~\bibnamefont {Wang}},
  \bibinfo {author} {\bibfnamefont {Z.-Q.}\ \bibnamefont {Xu}}, \bibinfo
  {author} {\bibfnamefont {J.}~\bibnamefont {Liu}}, \bibinfo {author}
  {\bibfnamefont {J.}~\bibnamefont {Song}}, \bibinfo {author} {\bibfnamefont
  {Y.}~\bibnamefont {Xue}}, \bibinfo {author} {\bibfnamefont {Z.}~\bibnamefont
  {Wang}}, \bibinfo {author} {\bibfnamefont {J.}~\bibnamefont {Zheng}},
  \bibinfo {author} {\bibfnamefont {L.}~\bibnamefont {Jiang}}, \bibinfo
  {author} {\bibfnamefont {C.}~\bibnamefont {Zheng}},  \emph {et~al.},\
  }\href@noop {} {\bibfield  {journal} {\bibinfo  {journal} {ACS nano}\
  }\textbf {\bibinfo {volume} {10}},\ \bibinfo {pages} {7031} (\bibinfo {year}
  {2016})}\BibitemShut {NoStop}%
\bibitem [{\citenamefont {Yuan}\ \emph {et~al.}(2016)\citenamefont {Yuan},
  \citenamefont {Wang}, \citenamefont {Shao}, \citenamefont {Lu}, \citenamefont
  {Li}, \citenamefont {Gruverman},\ and\ \citenamefont
  {Huang}}]{yuan2016electric}%
  \BibitemOpen
  \bibfield  {author} {\bibinfo {author} {\bibfnamefont {Y.}~\bibnamefont
  {Yuan}}, \bibinfo {author} {\bibfnamefont {Q.}~\bibnamefont {Wang}}, \bibinfo
  {author} {\bibfnamefont {Y.}~\bibnamefont {Shao}}, \bibinfo {author}
  {\bibfnamefont {H.}~\bibnamefont {Lu}}, \bibinfo {author} {\bibfnamefont
  {T.}~\bibnamefont {Li}}, \bibinfo {author} {\bibfnamefont {A.}~\bibnamefont
  {Gruverman}}, \ and\ \bibinfo {author} {\bibfnamefont {J.}~\bibnamefont
  {Huang}},\ }\href@noop {} {\bibfield  {journal} {\bibinfo  {journal}
  {Advanced Energy Materials}\ }\textbf {\bibinfo {volume} {6}},\ \bibinfo
  {pages} {1501803} (\bibinfo {year} {2016})}\BibitemShut {NoStop}%
\bibitem [{\citenamefont {Eames}\ \emph {et~al.}(2015)\citenamefont {Eames},
  \citenamefont {Frost}, \citenamefont {Barnes}, \citenamefont {O’regan},
  \citenamefont {Walsh},\ and\ \citenamefont {Islam}}]{eames2015ionic}%
  \BibitemOpen
  \bibfield  {author} {\bibinfo {author} {\bibfnamefont {C.}~\bibnamefont
  {Eames}}, \bibinfo {author} {\bibfnamefont {J.~M.}\ \bibnamefont {Frost}},
  \bibinfo {author} {\bibfnamefont {P.~R.}\ \bibnamefont {Barnes}}, \bibinfo
  {author} {\bibfnamefont {B.~C.}\ \bibnamefont {O’regan}}, \bibinfo {author}
  {\bibfnamefont {A.}~\bibnamefont {Walsh}}, \ and\ \bibinfo {author}
  {\bibfnamefont {M.~S.}\ \bibnamefont {Islam}},\ }\href@noop {} {\bibfield
  {journal} {\bibinfo  {journal} {Nature communications}\ }\textbf {\bibinfo
  {volume} {6}},\ \bibinfo {pages} {1} (\bibinfo {year} {2015})}\BibitemShut
  {NoStop}%
\bibitem [{\citenamefont {Barboni}\ and\ \citenamefont
  {De~Souza}(2018)}]{barboni2018thermodynamics}%
  \BibitemOpen
  \bibfield  {author} {\bibinfo {author} {\bibfnamefont {D.}~\bibnamefont
  {Barboni}}\ and\ \bibinfo {author} {\bibfnamefont {R.~A.}\ \bibnamefont
  {De~Souza}},\ }\href@noop {} {\bibfield  {journal} {\bibinfo  {journal}
  {Energy \& Environmental Science}\ }\textbf {\bibinfo {volume} {11}},\
  \bibinfo {pages} {3266} (\bibinfo {year} {2018})}\BibitemShut {NoStop}%
\bibitem [{\citenamefont {Courtier}\ \emph {et~al.}(2018)\citenamefont
  {Courtier}, \citenamefont {Richardson},\ and\ \citenamefont
  {Foster}}]{courtier2018fast}%
  \BibitemOpen
  \bibfield  {author} {\bibinfo {author} {\bibfnamefont {N.}~\bibnamefont
  {Courtier}}, \bibinfo {author} {\bibfnamefont {G.}~\bibnamefont
  {Richardson}}, \ and\ \bibinfo {author} {\bibfnamefont {J.~M.}\ \bibnamefont
  {Foster}},\ }\href@noop {} {\bibfield  {journal} {\bibinfo  {journal}
  {Applied Mathematical Modelling}\ }\textbf {\bibinfo {volume} {63}},\
  \bibinfo {pages} {329} (\bibinfo {year} {2018})}\BibitemShut {NoStop}%
\bibitem [{\citenamefont {Belova}\ and\ \citenamefont
  {Murch}(2004)}]{belova2004analysis}%
  \BibitemOpen
  \bibfield  {author} {\bibinfo {author} {\bibfnamefont {I.}~\bibnamefont
  {Belova}}\ and\ \bibinfo {author} {\bibfnamefont {G.}~\bibnamefont {Murch}},\
  }\href@noop {} {\bibfield  {journal} {\bibinfo  {journal} {Philosophical
  Magazine}\ }\textbf {\bibinfo {volume} {84}},\ \bibinfo {pages} {2139}
  (\bibinfo {year} {2004})}\BibitemShut {NoStop}%
\bibitem [{\citenamefont {Moleko}\ \emph {et~al.}(1989)\citenamefont {Moleko},
  \citenamefont {Allnatt},\ and\ \citenamefont {Allnatt}}]{moleko1989self}%
  \BibitemOpen
  \bibfield  {author} {\bibinfo {author} {\bibfnamefont {L.}~\bibnamefont
  {Moleko}}, \bibinfo {author} {\bibfnamefont {A.}~\bibnamefont {Allnatt}}, \
  and\ \bibinfo {author} {\bibfnamefont {E.}~\bibnamefont {Allnatt}},\
  }\href@noop {} {\bibfield  {journal} {\bibinfo  {journal} {Philosophical
  Magazine A}\ }\textbf {\bibinfo {volume} {59}},\ \bibinfo {pages} {141}
  (\bibinfo {year} {1989})}\BibitemShut {NoStop}%
\bibitem [{\citenamefont {Govinda}\ \emph {et~al.}(2017)\citenamefont
  {Govinda}, \citenamefont {Kore}, \citenamefont {Bokdam}, \citenamefont
  {Mahale}, \citenamefont {Kumar}, \citenamefont {Pal}, \citenamefont
  {Bhattacharyya}, \citenamefont {Lahnsteiner}, \citenamefont {Kresse},
  \citenamefont {Franchini} \emph {et~al.}}]{govinda2017behavior}%
  \BibitemOpen
  \bibfield  {author} {\bibinfo {author} {\bibfnamefont {S.}~\bibnamefont
  {Govinda}}, \bibinfo {author} {\bibfnamefont {B.~P.}\ \bibnamefont {Kore}},
  \bibinfo {author} {\bibfnamefont {M.}~\bibnamefont {Bokdam}}, \bibinfo
  {author} {\bibfnamefont {P.}~\bibnamefont {Mahale}}, \bibinfo {author}
  {\bibfnamefont {A.}~\bibnamefont {Kumar}}, \bibinfo {author} {\bibfnamefont
  {S.}~\bibnamefont {Pal}}, \bibinfo {author} {\bibfnamefont {B.}~\bibnamefont
  {Bhattacharyya}}, \bibinfo {author} {\bibfnamefont {J.}~\bibnamefont
  {Lahnsteiner}}, \bibinfo {author} {\bibfnamefont {G.}~\bibnamefont {Kresse}},
  \bibinfo {author} {\bibfnamefont {C.}~\bibnamefont {Franchini}},  \emph
  {et~al.},\ }\href@noop {} {\bibfield  {journal} {\bibinfo  {journal} {The
  journal of physical chemistry letters}\ }\textbf {\bibinfo {volume} {8}},\
  \bibinfo {pages} {4113} (\bibinfo {year} {2017})}\BibitemShut {NoStop}%
\end{thebibliography}

% Produces the bibliography via BibTeX.
\end{document}